\documentclass[aps,prl,twocolumn,showpacs,superscriptaddress]{revtex4}
\usepackage{amssymb}
\usepackage{textcomp}
\usepackage{graphicx} 
\usepackage{dcolumn}
\usepackage{bm}
\usepackage[dvips]{color}

\def\um{\mu\mbox{m}}

\def\Wcm2{\mbox{W cm}^{-2}}
\def\Wcmum2{\mbox{Wcm}^{-2}\mu\mbox{m}^{2}}

\def\cm3{\mbox{cm}^{-3}}

\definecolor{red}{rgb}{1,0,0}
\definecolor{blue}{rgb}{0,0,1}

\begin{document}

\title{Ion acceleration  in multispecies targets driven by intense laser radiation pressure}

\author{S.~Kar}\email{s.kar@qub.ac.uk}
\affiliation{Centre for Plasma Physics, School of Mathematics and Physics, Queen's University Belfast, BT7 1NN, UK}

\author{K.F.~Kakolee}
\affiliation{Centre for Plasma Physics, School of Mathematics and Physics, Queen's University Belfast, BT7 1NN, UK}

\author{B.~Qiao}
\affiliation{Centre for Plasma Physics, School of Mathematics and Physics, Queen's University Belfast, BT7 1NN, UK}
\affiliation{Centre for Energy Research, University of California San Diego, CA 92093-0417, USA}

\author{A.~Macchi}
\affiliation{Istituto Nazionale di Ottica, CNR, Pisa, Italy}
\affiliation{Department of Physics ``E. Fermi'', Largo B. Pontecorvo 3, 56127 Pisa, Italy}

\author{M.~Cerchez}
\affiliation{Institut f\"ur Laser-und Plasmaphysik, Heinrich-Heine-Universit\"at, D\"usseldorf, Germany}

\author{D.~Doria}
\affiliation{Centre for Plasma Physics, School of Mathematics and Physics, Queen's University Belfast, BT7 1NN, UK}

\author{M.~Geissler}
\affiliation{Centre for Plasma Physics, School of Mathematics and Physics, Queen's University Belfast, BT7 1NN, UK}

\author{P.~McKenna}
\affiliation{Department of Physics, SUPA, University of Strathclyde, Glasgow G4 0NG}

\author{D.~Neely}
\affiliation{Central Laser Facility, Rutherford Appleton Laboratory, Didcot, Oxfordshire, OX11 0QX, UK}

\author{J.~Osterholz}
\affiliation{Institut f\"ur Laser-und Plasmaphysik, Heinrich-Heine-Universit\"at, D\"usseldorf, Germany}

\author{R.~Prasad}
\affiliation{Centre for Plasma Physics, School of Mathematics and Physics, Queen's University Belfast, BT7 1NN, UK}

\author{K.~Quinn}
\affiliation{Centre for Plasma Physics, School of Mathematics and Physics, Queen's University Belfast, BT7 1NN, UK}

\author{B.~Ramakrishna}
\affiliation{Centre for Plasma Physics, School of Mathematics and Physics, Queen's University Belfast, BT7 1NN, UK}
\thanks{currently at Insitute of Radiation physics, HZDR, Germany}

\author{G.~Sarri}
\affiliation{Centre for Plasma Physics, School of Mathematics and Physics, Queen's University Belfast, BT7 1NN, UK}

\author{O.~Willi}
\affiliation{Institut f\"ur Laser-und Plasmaphysik, Heinrich-Heine-Universit\"at, D\"usseldorf, Germany}

\author{X.Y~Yuan}
\affiliation{Department of Physics, SUPA, University of Strathclyde, Glasgow G4 0NG}

\author{M.~Zepf}
\affiliation{Centre for Plasma Physics, School of Mathematics and Physics, Queen's University Belfast, BT7 1NN, UK}

\author{M.~Borghesi}
\affiliation{Centre for Plasma Physics, School of Mathematics and Physics, Queen's University Belfast, BT7 1NN, UK}

\date{\today}

\begin{abstract}
The acceleration of ions from ultra-thin foils has been investigated using 250 TW, sub-ps laser pulses, focused on target at intensities up to $3\times10^{20}~\Wcm2$. The ion spectra show the appearance of narrow band features for proton and Carbon peaked at higher energy (in the 5-10 MeV/nucleon range) and with significantly higher flux than previously reported. The spectral features, and their scaling with laser and target parameters, provide evidence of a multispecies scenario of Radiation Pressure Acceleration in the Light Sail mode, as confirmed by analytical estimates and 2D Particle In Cell simulations. The scaling indicates that monoenergetic peaks with more than 100 MeV/nucleon energies are obtainable with moderate improvements of the target and laser characteristics, which are within reach of ongoing  technical developments.

\end{abstract}

\pacs {}

\maketitle

Significant attention has been paid lately to laser-driven ion acceleration, which potentially offer a compact, cost-effective alternative to conventional sources, with potential applications in scientific, technological and healthcare applications~\cite{Borghesi_FST_2006}. Most experimental research so far has dealt with the Target Normal Sheath Acceleration (TNSA) mechanism~\cite{Borghesi_FST_2006}, where ions are accelerated by space charge fields set up by relativistic electrons at the target surfaces. TNSA ion beams have typically broadband spectrum, modest conversion efficiency  and large divergence, and reported scaling trends $E\propto I_{0}^{1/2}$ ~\cite{Borghesi_FST_2006}, (where $E$ and $I_{0}$ represent maximum proton energy and peak laser intensity) which however may not apply to  higher intensities regimes\cite{Robson_2007}. A different mechanism, Radiation Pressure Acceleration (RPA) ~\cite{Esirkepov_PRL_2004,Esirkepov_PRL_2006,Wilks_PRL_1992,Kar_PRL_2008, Robinson_PPCF_2009,Macchi_PRL_2009,Macchi_NJP_2010,Qiao_PRL_2009, Qiao_PRL_2010,Qiao_POP_2011,Robinson_NJP_2008,Henig_PRL_2009, Klimo_PRE_2008,Yan_PRL_2008} is currently attracting a substantial amount of experimental and theoretical attention  due to the predicted superior scaling in terms of ion energy and laser-ion conversion efficiency. In this prospective, the so called 'Light Sail'(LS)~\cite{Macchi_PRL_2009,Macchi_NJP_2010,Robinson_NJP_2008,Qiao_PRL_2009,Qiao_PRL_2010,Qiao_POP_2011,Henig_PRL_2009,Klimo_PRE_2008,Yan_PRL_2008} scheme, where, in a sufficiently thin foil, the whole laser-irradiated area is detached and pushed forward by the Radiation Pressure, is particularly promising. 

Among the attractive features of the  LS mechanism are a favorable scaling with the laser fluence ~\cite{Macchi_PRL_2009,Macchi_NJP_2010,Robinson_NJP_2008}, natively narrow energy bandwidth, a reduced divergence, and a similarly efficient acceleration for both protons and higher mass ions, as predicted by numerous computational and analytical studies. Experimental evidence of RPA-LS is however scarce, as TNSA dominates in standard interaction conditions. Henig~\textit{et. al.}~\cite{Henig_PRL_2009} have reported on spectral features, which have been associated to RPA effects - namely a polarization-dependent modification of the carbon ion spectral profile at moderate ion energies. 

This Letter presents experimental evidence of narrow band spectral features emerging from thin foil irradiation by sub-PW laser pulses. In particular, Carbon ion peaks of up to $\sim7$ MeV/nucleon (cut-off energy in excess of 10 MeV/nucleon) are produced for the first time, with nearly an order of magnitude higher particle flux than previously reported~\cite{Henig_PRL_2009,Hegelich_Nature_2006,Jung_PRL_2011}. The spectral features, and their scaling with the laser and target parameters, point to a multispecies scenario of LS acceleration as described in~\cite{Yu_PRL_2010,Qiao_PRL_2010}. The possibility of achieving spectral peaks beyond 100 MeV/nucleon, a key requirement for hadron therapy~\cite{Amaldi_NP_1999}, by tuning currently achievable laser and target parameters is discussed on the basis of the experimental scaling, and supported by 2D Particle in Cell (PIC) simulations. 

\begin{figure*}
\begin{center}
\includegraphics[angle=0,width=0.95\textwidth]{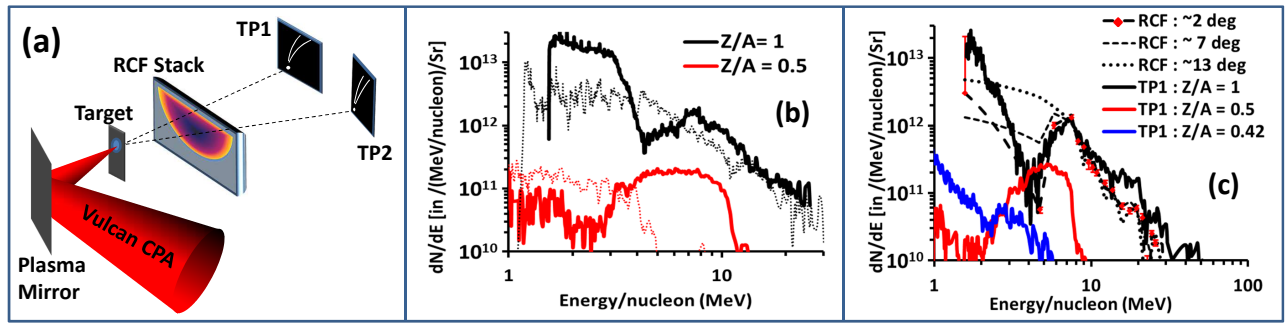}
\end{center}
\caption{\textbf{(a)} Schematic of the experimental setup. \textbf{(b)} Ion spectra obtained from 100 nm Cu target irradiated by a LP laser pulse at $I_0=\sim3\times10^{20}~\Wcm2$. Different line colors correspond to different ion species (see the figure legend), solid and dotted lines represent spectra obtained on TP1 and TP2 respectively. \textbf{(c)} Ion spectra obtained on TP1 from 50 nm Cu target irradiated by a CP pulse at $I_0\simeq1.25\times10^{20}~\Wcm2$. The proton spectra obtained with RCF, in the same shot, at different positions in the beam, corresponding to $\sim2^{o}$, $\sim7^{o}$ and $\sim13^{o}$ from the laser axis are also shown (dashed lines).}\label{setup}.
\end{figure*}

The experiment was carried out employing the PW arm of the VULCAN laser system at the Rutherford Appleton Laboratory, STFC, UK. A schematic of the experimental setup is shown in the Fig.~\ref{setup}(a). The laser delivered $\sim$200 J (varying from 175 J to 225 J in different shots) energy on target in pulses of  700 fs - 900 fs FWHM duration after being reflected off a Plasma Mirror (PM), resulting in intensity contrast ratio of $10^9$ between the main pulse and the ns long amplified spontaneous emission (ASE). The laser was focussed on targets at normal incidence angle by an $f/3$ off axis parabolic mirror. The laser peak intensity on the target was varied from $I_{0}= 5\times10^{19}~\Wcm2$ to $3\times10^{20}~\Wcm2$ by varying the focal spot size of the laser on target. The polarisation of the laser on the target was varied from Linear(LP) ($\varepsilon=0$), to elliptical (EP) ($\varepsilon \sim 0.47\pm0.02$) and nearly circular (CP) ($\varepsilon \sim 1.14\pm0.04$) by employing a zero order quarter wave plate placed between the focussing optics and the PM. Here, $\varepsilon$ is the ratio between the laser electric field amplitudes along the vertical and horizontal axes, as determined by the combined effect of the wave plate and of reflection from the PM. Targets of different composition and thickness were irradiated. Ion spectra produced by the interaction were diagnosed by two thompson parabola (TP) spectrometers (with acceptance cone of 19.8 nSr) as shown in the Fig.~\ref{setup}(a) - TP1 placed along the laser axis and TP2 at $13\pm 2$ degrees off axis. The energy-resolved spatial profile of  the bottom half of the ion beam was recorded by stacks of radiochromic films (RCF). The image plate detectors used in the TPs were cross-calibrated with CR39 solid state nuclear track detector for particle number density~\cite{image_plate_calibration}. Similarly, the RCF dose response  was absolutely calibrated~\cite{RCF_calibration}. 

While exponential spectra were always produced from  5-10 $\um$ thick foil targets, narrow-band features in proton and heavier ion spectra were observed from sub-$\um$ thick targets irradiated with high intensities.  For example, the spectra in Fig.~\ref{setup}(b), obtained from a 100 nm thick Cu target irradiated by a LP (p-polarization) laser pulse at peak intensity of $\sim3\times10^{20}~\Wcm2$, shows narrow-band peaked features in the proton (charge to mass ratio $Z/A=1$) and carbon ($Z/A=0.5$) spectra clearly separated from a lower energy component (as usual in standard interaction conditions, hydrogen and Carbon ions observed in the spectrum originate from surface contaminant layers). The lower $Z/A$ ion species had broad, exponential-like spectra similar to that shown in Fig.1 (c) for $Z/A \simeq 0.42$. Such charge state is likely to correspond to partially ionized Cu, as, for instance, a close inspection suggests a fine structure within the track possibly corresponding to closely spaced values of Z/A. The peaks in ion spectra appear to be ordered, with the proton peak at slightly higher energy than the C peak. It is also interesting to note the ion flux at the peaked features, which is nearly an order of magnitude higher than previously reported~\cite{Henig_PRL_2009,Hegelich_Nature_2006,Jung_PRL_2011}.  

The narrow-band spectral features observed in carbon spectra along the laser axis are not observed with the off axis detector TP2, as shown in Fig.~\ref{setup}(b), indicating that they are confined within a cone of half-aperture less than $13\pm 2~\deg$. 
This indication is corroborated by RCF stack data taken simultaneously to the spectral measurement with the TPs (see Fig.~\ref{setup}(c)). The stopping range of carbon is significantly shorter than for protons at the same energy/nucleon: for example, the stopping range of 5 MeV/nucleon proton and carbon in the RCF stack, protected by a 30 $\um$ Al foil, are $\sim240\um$ and $\sim60\um$ respectively. For this reason, the deposited dose in the RCFs in the stack can be assumed to be primarily due to the protons. By deconvolving the proton spectrum from the RCFs, the narrow band  feature in the proton spectrum observed in the TP1 was reproduced (dashed line in Fig.~\ref{setup}(c)) for a defined region in the RCF corresponding to a half cone beam divergence of $\sim10^{o}$. The proton spectrum gradually becomes exponential (filling the dip in the spectrum at the low energy side) as one moves farther from the laser axis. Such exponential spectra are typical of TNSA acceleration caused by the extended sheath produced by hot electrons at the target rear surface~\cite{Borghesi_FST_2006}. Assuming that the divergence of the narrow band feature in the proton beam and in the carbon ion beam are comparable, the conversion efficiency into this component can be estimated as $\sim1\%$, which is significantly higher than reported in~\cite{Jung_PRL_2011} and comparable to ref.\cite{Henig_PRL_2009}. 

\begin{figure}
\begin{center}
\includegraphics[angle=-90,width=0.47\textwidth]{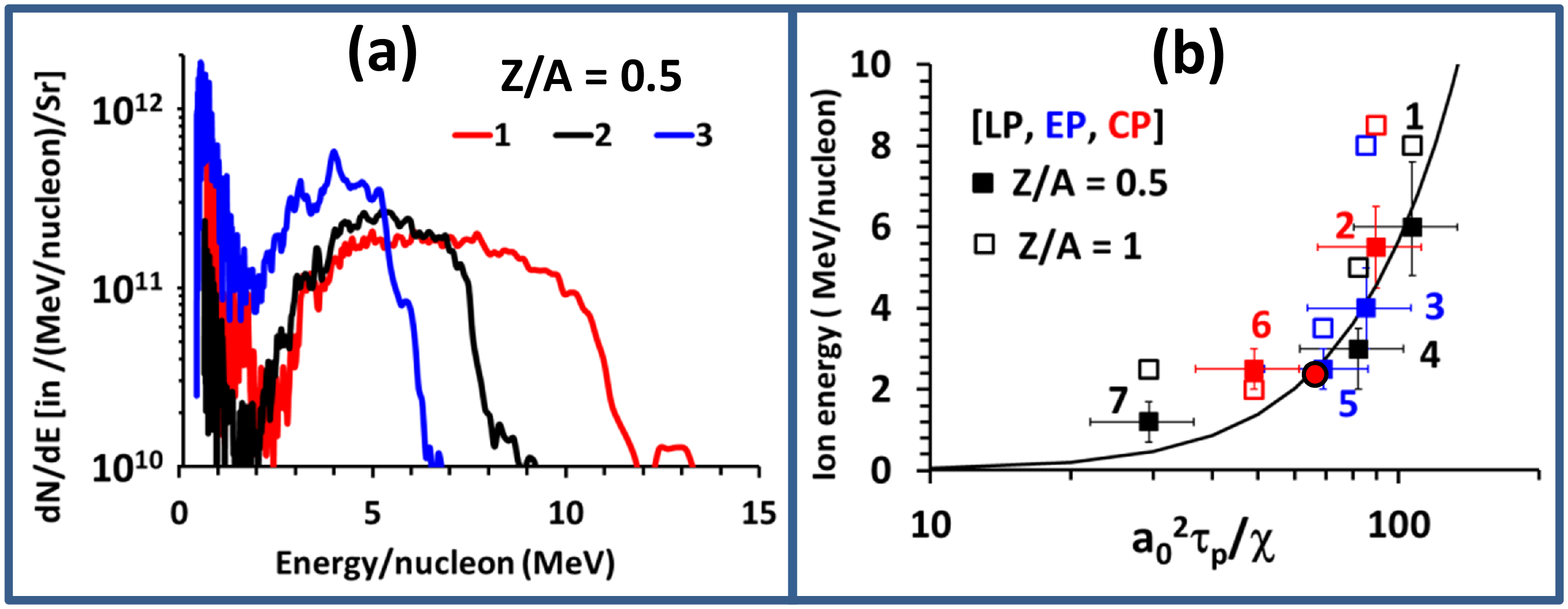}
\end{center}
\caption{\textbf{(a)} Graph showing comparison between five ion ($Z/A$ = 0.5) spectra, where the spectral peaks are plotted in \textbf{(b)} as a function of $a_0^2\tau_{p}/\chi$. The experimental parameter set [$a_0$, target material, target thickness($\um$)] for the data points 1-7 are [15.5, Cu, 0.1], [10, Cu, 0.05], [13.8, Cu, 0.1], [7.5, Al, 0.1], [6.9, Al, 0.1], [13.6, Al, 0.5] and [14.1, Al, 0.8] respectively. Black solid line in (b) represents ion energy estimated from the analytical modeling described in the text. The red circle represent the data shown by Henig~\textit{et. al.} in Ref.~\cite{Henig_PRL_2009}.}\label{carbon_bunch}
\end{figure}

The appearance and position of distinct peaks in the ion spectrum could be controlled by varying laser and target parameters as shown in Fig.~\ref{carbon_bunch}(a) and (b). Peaks were only observed in the limit of thin foils and high intensity (Fig.~\ref{carbon_bunch}(a)) with the peaks shifting towards high energy as either the intensity was increased or the target thickness was reduced (Fig.~\ref{carbon_bunch}(b))

In order to assess the possible influence of Radiation Pressure effects on the spectral profiles observed, a simple analytical model was developed taking into account the Hole Boring (HB) and LS phases of RPA mechanism. Due to the extreme radiation pressure (of the order of hundreds of GBar) exerted on the irradiated  target surface, ions are swept forward by directional momentum transfer~\cite{Wilks_PRL_1992,Kar_PRL_2008,Robinson_PPCF_2009}. Consequently, the laser pulse bores through the target, in the HB phase, with the ion front velocity (in units of $c$) given by~\cite{Wilks_PRL_1992,Macchi_PRL_2005,Robinson_PPCF_2009} 
\begin{equation}\label{EQ-HB}
\beta=\sqrt{\frac{I_0(t)}{\rho c^3}}=
\sqrt{\frac{m_{e}}{m_{p}}\frac{a^{2}(t)}{\rho^{\prime}}} \, ,
\end{equation}
where $a(t)=0.85\sqrt{I_0(t)\lambda^{2}/10^{18}~\Wcm2}$, $\rho^{\prime}=\rho/m_{p}n_{c}$ with $\rho$ the mass density, $m_{p}$ the proton mass and $n_{c}$ is the critical electron density, and $I_0(t)$ is the laser intensity on as a function of time. (Relativistic effects \cite{Robinson_PPCF_2009} are neglected here for simplicity, consistently with the values of $\beta\ll 1$ for our parameters.)
If the target is sufficiently thin, the hole boring front  will tend to reach the target rear surface at a time $t_{HB}<t_p$(laser pulse duration), i.e. before the end of the laser pulse. As the thickness of the compressed layer becomes comparable or less than the evanescence length of the ponderomotive force, the whole layer can be cyclically accelerated with high efficiency for the remainder of the laser pulse duration($t_{LS}$). In this scenario of whole foil acceleration (LS regime) the equations of motion of the foil can be expressed as ~\cite{Macchi_PRL_2009,Macchi_NJP_2010,Qiao_PRL_2009,Qiao_PRL_2010,Qiao_POP_2011}
\begin{equation}\label{EQ-LS}  
\gamma^{3}\frac{d\beta}{d\tau}=\frac{2m_{e}}{m_{p}}\frac{1-\beta}{1+\beta}
\frac{a^{2}(\tau-x)}{\chi}R \, ,\qquad 
\frac{dx}{d\tau}=\beta \, ,
\end{equation}
where $\tau=ct/\lambda$,
$\gamma=(1-\beta^{2})^{-1/2}$ and $\chi=\rho^{\prime}l/\lambda$ for a target of thickness $l$, under the approximation that the areal density of the compressed layer is same as that of the target before the interaction.
The reflectivity $R=R(\beta,\chi,a)$ may be estimated using a simple model of relativistically induced self-transparency \cite{Macchi_NJP_2010} and the
system of Eqs.(\ref{EQ-LS}) can be integrated numerically. However, some insight may already be obtained analytically.
Assuming $R \simeq 1$, Eq.(\ref{EQ-LS}) may be integrated to yield the final velocity of the foil $\beta_f$ and the corresponding energy per nucleon $E_{ion}=(\gamma_f-1)m_{p}c^{2}$ as a function of the dimensionless fluence parameter ${\cal E}=(\rho l c^2)^{-1}\int_{t_0}^{\infty}I_0(t)dt$ 
and of the initial velocity $\beta(t_0)=\beta_0$ \cite{Simmons_AJP_1993}:
\begin{equation}
\label{LS_scaling}
\frac{E_{ion}}{m_{p}c^{2}}=\frac{({\cal E}+Z_0)^2}{2({\cal E}+Z_0+1)} \, , \qquad
Z_0=\left(\frac{1+\beta_0}{1-\beta_0}\right)^{1/2}-1 \, .
\end{equation}
To match the initial HB stage with the later LS one, we take $t_0=t_{HB}$ and, from Eq.(\ref{EQ-HB}), $\beta_0=\sqrt{I(t_0)/\rho c^3}$. For thin targets and high intensities, $t_{HB}\ll t_p \simeq t_{LS}$ and the LS stage will dominate also because of higher efficiency with respect to HB, thus ${\cal E}\gg Z_0$. In such limit we may estimate ${\cal E} \simeq a_{0}^{2}\tau_{LS}/\chi \simeq a_{0}^{2}\tau_{p}/\chi$ where  $\tau_{p}=ct_{p}/\lambda$ and $a_0^2\tau_{p}=\int^\infty_{-\infty}{a^2(\tau)~d\tau}$, and $E_{ion}$ will scale as $(a_{0}^{2}\tau_{p}/\chi)^{\alpha}$ with $\alpha=2$ for ${\cal E}\ll 1$ and  $\alpha=1$ for ${\cal E}\gg 1$, i.e. in the ultra-relativistic case~\cite{Esirkepov_PRL_2004}.

On the basis of the above equations, we can estimate whether the peaks observed, and their scaling with the target and laser parameters are consistent with the expectations for LS. The switch-on time for the LS phase is taken as the time at which the compressed ion front reaches the target rear surface as a result of the HB process. In this simple rigid model, the degree of ionisation of target ions and the number of different ion species present in the compressed layer are irrelevant. Therefore, without any loss of generality, we have  ignored the target contaminant layers, which have typically  few nm thickness and significantly lower density than the metal targets used in our case. As shown in the Fig.~\ref{carbon_bunch}(b), the ion energy estimated by the model agrees well with the data, within the experimental errors. As expected for the non-relativistic case, the ion energy is seen to scale as $(a_{0}^{2}\tau_{p}/\chi)^{2}$. For comparison, the only published data point for carbon peaks attributed to RPA-LS~\cite{Henig_PRL_2009} is also shown in the graph, which is in substantial agreement with the calculated scaling. For the data points shown in the Fig.~\ref{carbon_bunch}(b), the hole boring phase ends significantly before the peak of the laser pulse and the non-linear reflectivity remains at unity for the duration of the pulse.

The experimental features described above are observed virtually independent of the incident laser polarization. This is not surprising since the cycle-averaged light pressure is the same for CP and LP. While the early simulations of RPA were carried out employing LP pulses at extreme intensities ~\cite{Esirkepov_PRL_2004,Esirkepov_PRL_2006}, the attention of theoretical work over the past few years has focused on the use of circular polarization as a means to minimize effects of  electron heating and sheath acceleration, and achieve a stable LS drive. Multi-dimensional simulations show that the stability of the LS acceleration is dominated by factors such as the shape of the radial laser intensity profile~\cite{Robinson_NJP_2008,Macchi_NJP_2010} and achieving a smooth  transition from the HB phase to the LS phase at the early stages of the acceleration process~\cite{Bin_arxiv}. The polarization, while still affecting the outcome, plays a sub-ordinate role in this case. One should note that, in the experiment the effect of changing the polarization of the laser pulse may be weakened by several factors, such as the polarisation-dependent reflectivity of the plasma mirror (so that circular polarization actually corresponds to an elliptical polarisation on target with $\epsilon\sim$ 1.14), the tight $f/3$ focusing resulting a spatially Gaussian intensity profile on target, and by the relatively long duration of the laser pulse. The latter two factors lead to a significant deformation of the ultra-thin target foil during the acceleration so that normal incidence is not strictly preserved locally. All these factors enhance electron heating and reduce any polarisation dependance. Furthermore, recent work  by Qiao et al.~\cite{Bin_arxiv} has highlighted that efficient LS drive can be achieved with LP pulses at presently achievable intensities, in a regime where RPA and TNSA co-exist. In this framework, sheath-field effects may account for the observed separation of the C and proton spectral peaks [see Fig.~\ref{setup}(b),(c)] (as in an ideal  LS scenario all ions in the sail should reach the same energy per nucleon). After the LS stage, the sheath field can in principle further accelerate the ions, with the  protons gaining more energy/nucleon than Carbon ions due to their higher $Z/A$ . This last acceleration stage may preserve or even enforce the peaked distribution as it typically happens in a multispecies expansion \cite[and references therein]{Tikhonchuk_PPCF_2005}.

This scenario is confirmed by 2D PIC simulation employing the "ILLUMINATION" code~\cite{illumination}. Due to limitation in computational resources which prevent $\sim$ps scale simulation of long pulse interaction, we scaled down the laser and target parameters in such a way that both the dimensionless quantities $I_0/\rho c^3$ and $(a_{0}^{2}\tau_{p}/\chi)$ remain the same as the case shown in the Fig.~\ref{setup}(b). A LP laser pulse with $\lambda=1.0\mu\textrm{m}$, $I_0=7.5\times10^{19}\Wcm2$ ($a_0=7.75$), gaussian spatial profile of $5\um$ radius and gaussian temporal profile of 50 fs FWHM was used.  A copper target was used with electron density of $384.3n_c$ and thickness of $l_0=10$ nm, covered on both side with 3~nm, $48.0n_c$ layer of hydrocarbon contaminants. The simulation box was $20~\um$ long and $24.576~\um$ wide with 1 nm $\times$ 8 nm cell size in order to assure resolution for the contaminant layers. 72 particles per cell for $\textrm{Cu}^{28+}$ species and 36 particles per cell for $\textrm{H}^+$ and $\textrm{C}^{6+}$ in the contaminant layers were used.  

\begin{figure}[h!]
\begin{center}
\includegraphics[angle=0,width=0.47\textwidth]{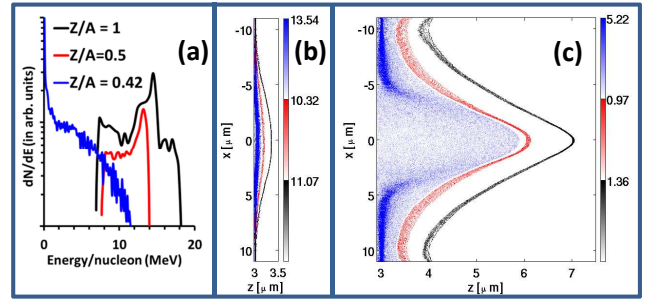}
\end{center}
\caption{\textbf{(a)} Ion spectra obtained from 2D multilayered PIC simulation for the data point '1' in Fig.~\ref{carbon_bunch}(a), of which the experimental spectra is shown in the Fig.~\ref{setup}(b). \textbf{(b)} and \textbf{(c)} shows 2D density profiles of ions (black-proton, red-carbon and blue-Copper) at 66 fs and 165 fs, respectively. Each colormap ranges from zero to the value mentioned in the graph.}\label{simulation}
\end{figure}

As expected, the simulation shows that TNSA and HB takes place with the rising intensity of the laser pulse, accelerating ions from rear and front surfaces respectively. After the ions from the front surface pile up at the target rear surface forming a compressed layer, ions are accelerated predominantly by the LS mechanism resulting in the ion spectrum shown in the Fig.~\ref{simulation}(a), which reproduces the main features observed in the experimental data. The target remains highly reflective during the whole duration of the simulation and the carbon ions are accelerated in a snowplough fashion maintaining a dense layer as shown in the Fig.~\ref{simulation}(b) and (c). It is interesting to note the debunching of the Cu ions after the laser pulse~\cite{Qiao_POP_2011}, as shown in the Fig.~\ref{simulation}(c), producing the exponential spectrum profile of  Fig.~\ref{simulation}(a), which resembles closely the experimental data. This is consistent with the scenario highlighted in~\cite{Qiao_PRL_2010}, where the heavier species in a multispecies target undergo decompression and Coulomb explosion, while the lighter species are stabilized by the excess electrons released by the heavier species.

\begin{figure}[t]
\begin{center}
\includegraphics[angle=-90,width=0.47\textwidth]{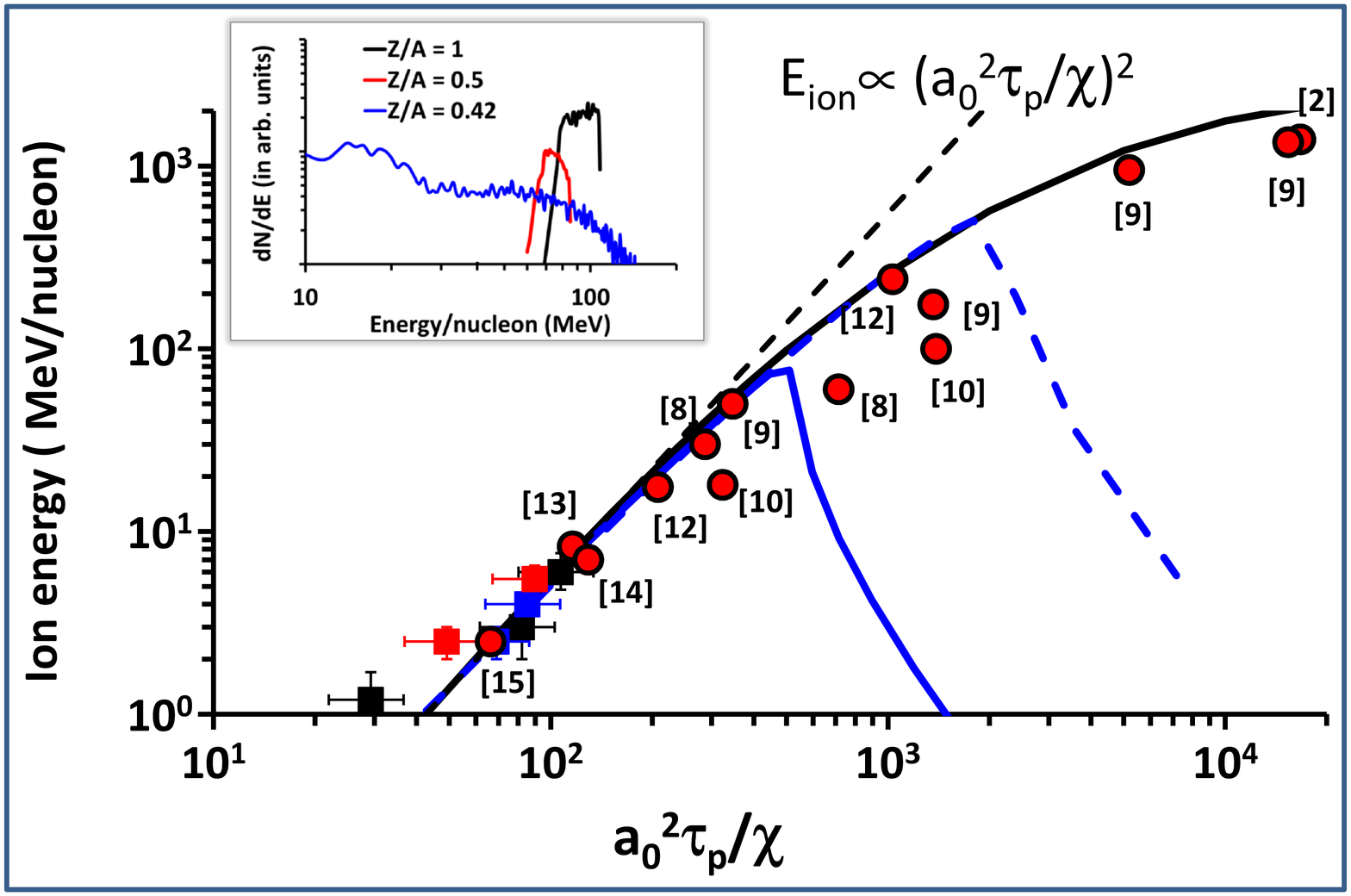}
\end{center}
\caption{Ion energy scaling (black solid line) as shown in the Fig.~\ref{carbon_bunch}(b), extrapolated to higher $a_{0}^{2}\tau_{p}/\chi$, 
assuming $R=1$.
Square points are the experimental data shown in the same figure. The red circles represent the results reported in the literature, as labeled, by multispecies PIC simulations for stable LS acceleration. Dashed black line shows trend for a $E_{ion}\propto(a_{0}^{2}\tau_{p}/\chi)^2$ scaling valid for non-relativistic ion energies. Solid and dashed blue lines plots expected ion energy (according to the rigid model) by varying $\chi$ for 45 fs FWHM laser at $I_{0}=5\times10^{20}\Wcm2$ and 450 fs FWHM laser at $I_{0}=5\times10^{19}\Wcm2$ respectively. The insert shows the spectra obtained from a PIC simulation identical to the run shown in the Fig.~\ref{carbon_bunch}(c), except the laser fluence increased by a factor of 2 and target density decreased by a factor of 2.5.}
\label{scaling}
\end{figure}

The experimental data points in  Fig.~\ref{carbon_bunch}(b) show that the ion energy scales with the parameter $(a_{0}^{2}\tau_{p}/\chi)$ to the power 2, as expected for a fully reflecting target, and in the limit $\beta\ll 1$. As reported by several groups via extensive 2D and 3D simulations (see Fig.~\ref{scaling}), the ion energy can therefore be enhanced by increasing laser fluence and decreasing the target areal density. However, in doing this,  one needs to avoid  self-induced transparency, as this terminates efficient LS (see the blue lines in the Fig.~\ref{scaling}) and leads to a dramatical reduction of the ion energy.

Although the  ion energies  achieved here  ($\textgreater$10 MeV/nucleon) are encouraging, producing peaks at more than 100 MeV/nucleon would be a crucial milestone in view of applications. As shown in the insert in the Fig.~\ref{scaling}, 2D PIC simulations predict that  100 MeV/nucleon ions in a narrow energy bandwidth can be reached by, for example, increasing the laser fluence by a factor of 2 and decreasing the target density by a factor of 2.5, compared to the case shown in the Fig.~\ref{setup}(b). This seems achievable given current developments in laser technology.

In conclusion, we have reported on the observation of narrow band features in the spectra of laser-accelerated ions, which appear to be consistent with Radiation pressure acceleration, in a regime where LS overcomes sheath acceleration. The observed ion  beams show unique properties in terms of ion energy, divergence and fast energy scaling, and offer high promise for further progress in the near future.

\begin{acknowledgments}
Authors acknowledge funding from EPSRC [EP/E035728/1 - LIBRA consortium, EP/J002550/1 - Career Acceleration Fellowship by SK and EP/E048668/1], Leverhulme Trust Fellowship (ECF-2011-383) hold GS, MIUR (Italy) via the FIRB project ``SULDIS'', and DFG programes TR18 and GK1203. Authors also acknowledge support from target fabrication group and e-Science facility of RAL-STFC.
\end{acknowledgments}

\end{document}